\documentclass[10pt,a4paper]{article}
\usepackage[utf8]{inputenc}
\usepackage[english]{babel}

\usepackage{amsmath,amsfonts,amssymb}
\usepackage{titling,authblk}
\usepackage[left=2cm,right=2cm]{geometry}

\usepackage{feynmp}
\RequirePackage[colorlinks,citecolor=blue,urlcolor=blue,linkcolor=blue]{hyperref}

\title{Graviton Mixing}

\author[1,2]{\href{https://orcid.org/0000-0001-7099-0861}{Boris Latosh} \footnote{\href{mailto:latosh@theor.jinr.ru}{latosh@theor.jinr.ru}}}
\affil[1]{Bogoliubov Laboratory for Theoretical Physics, JINR, Dubna 141980, Russia;}
\affil[2]{Dubna State University, Universitetskaya str. 19, Dubna 141982, Russia}

\begin{document}

\maketitle

\begin{abstract}
  Mechanism mixing graviton spin states is defined. The mixing appears naturally due to loop corrections. Its influence on amplitudes involving matter states is shown, implications for empirical data are discussed. It is argued that the mixing is one of the reasons behind an inability to defined the universal running of the Newton constant.
\end{abstract}

\section{Introduction}

Quantum general relativity is known to experience problems with renormalizability. At the one-loop level pure general relativity is renormalizable only on-shell \cite{tHooft:1974toh}. When matter is added to the model, then even on-shell renormalization does not take place \cite{tHooft:1974toh}. At the two-loop level even pure general relativity is completely non-renormalizable \cite{Goroff:1985th}. Because of this it can hardly be considered as a fundamental theory.

Effective field theory technique provides a way to account for quantum corrections to general relativity consistently \cite{Burgess:2003jk,Donoghue:2012zc,Donoghue:1994dn} and avoid problems associated with non-renor\-malizability. The technique is based on a premise that general relativity describes gravitational interaction at some energy scale $\mu$ below the Planck scale. From the normalization scale $\mu$ the theory is extended to the low energy regime via standard loop corrections. Applicability of the theory is limited only to energies below the normalization scale and it is assumed that all divergences in loop corrections can be renormalized within the complete theory.

The effective theory for general relativity allows one to obtain some verifiable predictions, for instance to recover the low energy effective action \cite{Donoghue:1994dn,Burgess:2003jk}, to obtain corrections to the Newton law \cite{BjerrumBohr:2002kt,Bjerrum-Bohr:2016hpa}, etc \cite{Calmet:2018qwg,Calmet:2018uub}.

In this paper we highlight a phenomenon that we call ``graviton mixing''. It is similar to the fermion mixing in the standard model. In the electroweak sector of the standard model neutrino free states which are eigenstates of the mass operator $(\nu_1,\nu_2,\nu_3)$ are given as a superposition of eigenstates of an interaction operator $(\nu_e,\nu_\mu,\nu_\tau)$. Their relation is described by the PMNS matrix \cite{Pontecorvo:1957qd,Maki:1962mu,Tanabashi:2018oca}. To put it otherwise, fermion mixing takes place, when a superposition of fermion state with well-defined mass is coupled to gauge bosons \cite{Tanabashi:2018oca}.

A similar phenomenon takes place in gravity. General relativity contains massless spin-2 degree of freedom on shell, while off-shell it receives an additional massless scalar degree of freedom \cite{Fierz:1939ix,Hinterbichler:2011tt,RN7}. In the classical theory these degrees of freedom are coupled to an energy-momentum source $T^{\mu\nu}$ in the same way:
\begin{align}
  T^{\mu\nu} h_{\mu\nu} = T^{\mu\nu} h_{\mu\nu}^{(s=2)} + T^{\mu\nu} h_{\mu\nu}^{(s=0)}.
\end{align}
Within an effective theory this is not so, as the correspondent couplings are modified due to quantum effects. In full analogy with the case of the standard model one should introduce a mixing matrix $\mathcal{M}_{\mu\nu\alpha\beta}$ that couples a linear combination of spin-2 and spin-0 states to an energy-momentum source:
\begin{align}
  &T^{\mu\nu} h_{\mu\nu} \to T^{\mu\nu} \mathcal{M}_{\mu\nu\alpha\beta} h^{\alpha\beta}= \nonumber \\
  &=T^{\mu\nu} ~\left[ M_{\mu\nu\alpha\beta} h^{(s=2)}{}^{\alpha\beta} + M_{\mu\nu\alpha\beta} h^{(s=0)}{}^{\alpha\beta}\right].
\end{align}

The main focus of the paper is to draw an attention to the graviton mixing and to highlight its role within effective gravity. In the next section we discuss a motivation behind the graviton mixing in more details and provide a way to construct a graviton mixing matrix. Then we show that the mixing has a non-trivial influence on processes involving matter. Namely, the structure of virtual graviton exchange processes is strongly affected by the mixing. We conclude with a discussion of the role of graviton mixing.

\section{Graviton Mixing}

Within general relativity an interaction between a week gravitational field $h_{\mu\nu}$ and an energy-momentum source is given by the following Lagrangian:
\begin{align}
  \mathcal{L}_\text{int} = T^{\mu\nu} ~I_{\mu\nu\alpha\beta} h^{\alpha\beta}.
\end{align}
Here $I_{\mu\nu\alpha\beta}$ is the gauge-invariant generalization of a rank-$2$ unit tensor
\begin{align}
  I_{\mu\nu\alpha\beta} \overset{\text{def}}{=} \cfrac12 \left( \theta_{\mu\alpha}\theta_{\nu\beta}+\theta_{\mu\beta} \theta_{\nu\alpha} \right) .
\end{align}
And the standard projectors $\theta_{\mu\nu} = \eta_{\mu\nu} - k_\mu k_\nu/k^2$ are used. The interaction contains spin-$2$ and spin-$0$ parts:
\begin{align}\label{interaction_gauge-invariant_form}
  L_\text{int}&= T^{\mu\nu} P^2_{\mu\nu\alpha\beta} h^{\alpha\beta} + T^{\mu\nu} P^0_{\mu\nu\alpha\beta} h^{\alpha\beta} \nonumber \\
  &=T^{\mu\nu} h^{(s=2)}_{\mu\nu} + T^{\mu\nu} h^{(s=0)}_{\mu\nu}.
\end{align}
Where $P^2$ and $P^0$ are Nieuwenhuizen operators \cite{VanNieuwenhuizen:1973fi,Accioly:2000nm}:
\begin{align}
  \begin{cases}
    P^2_{\mu\nu\alpha\beta} &\overset{\text{def}}{=} \cfrac12 \left( \theta_{\mu\alpha} \theta_{\nu\beta} + \theta_{\mu\beta} \theta_{\nu\alpha}\right) - \cfrac13 ~ \theta_{\mu\nu} \theta_{\alpha\beta}, \\
    P^0_{\mu\nu\alpha\beta} &\overset{\text{def}}{=} \cfrac13 ~ \theta_{\mu\nu} \theta_{\alpha\beta}.
  \end{cases}
\end{align}
They are orthogonal projectors on spin-2 and spin-0 states:
\begin{align}
  \begin{cases}
    P^2_{\mu\nu\rho\sigma} P^2{}^{\rho\sigma}{}_{\alpha\beta} &= P^2_{\mu\nu\alpha\beta}, \\
    P^0_{\mu\nu\rho\sigma} P^0{}^{\rho\sigma}{}_{\alpha\beta} &= P^0_{\mu\nu\alpha\beta}, \\
    P^2_{\mu\nu\rho\sigma} P^0{}^{\rho\sigma}{}_{\alpha\beta} &= 0 .
  \end{cases}
\end{align}
Thus, in general relativity spin-$2$ and spin-$0$ graviton states share the matter coupling constant.

Within a more general setup of effective theory this feature can hardly holds. Even if general relativity does describe gravity at some energy scale $\mu$ below the Planck scale, then loop corrections can still induce a non-minimal coupling between graviton spin states and matter. In the most general case one can account for such effects by introducing the graviton mixing matrix:
\begin{align}
  \mathcal{M}_{\mu\nu\alpha\beta} = I_{\mu\nu\alpha\beta} + A P^2_{\mu\nu\alpha\beta} + B P^0_{\mu\nu\alpha\beta}
\end{align}
with $A$ and $B$ being mixing parameters. The matrix redefines the form of the interaction between gravity and matter:
\begin{align}
  \mathcal{L}_\text{int}\to T^{\mu\nu} \mathcal{M}_{\mu\nu\alpha\beta} h^{\alpha\beta}.
\end{align}
A few comment on the mixing are due.

Firstly, the mixing parameters $A$ and $B$ should be given in terms of momentum expansions. Loop corrections can be consistently taken into account within effective field theory \cite{Donoghue:1994dn,Burgess:2003jk}. The form of the mixing parameters can be recovered via dimension reasoning:
\begin{align}\label{mixing_parameters_expansion}
  \begin{cases}
    A (\kappa^2\square) = a_1 \kappa^2 \square + a_2 (\kappa^2 \square)^2 + \cdots \\
    B (\kappa^2\square)= b_1 \kappa^2 \square + b_2 (\kappa^2 \square)^2 + \cdots
  \end{cases}.
\end{align}
These expansions are due, as the $n$-loop graviton correction is suppressed by the factor $\kappa^{2n}$. Thus, constants $a_1$, $b_1$ contain data on one-loop corrections, $a_2$ and $b_2$ describe two-loop corrections, etc. The explicit values of these constants cannot be evaluated within the effective theory, but can be calculated within quantum gravity models.

Secondly, mixing parameters are defined by the particle content of the effective theory. Alongside graviton loop corrections the energy-momentum tensor is also affected by matter loop corrections. It is safe to consider only renormalizable interactions with dimensionless couplings. Corrections from such matter loops do not influence the structure of the momentum expansions \eqref{mixing_parameters_expansion} and only contribute to coefficients $a_i$, $b_i$. Cases of non-renormalizable interactions and interactions with dimensionful couplings require a special treatment and lie beyond the scope of this paper.

Summarizing all of the above, within an effective theory for general relativity the graviton mixing appears naturally due to loop corrections. The mixing parameters are defined by the structure of the underlying fundamental theory, but their form can be recovered via dimension reasoning. But most importantly, the mixing must be taken into account for processes involving matter states.

The importance of the graviton mixing can be easily illustrated via a simple example of a virtual graviton exchange. To proceed with the goal one should use the gauge-invariant part of the graviton propagator \cite{Accioly:2000nm}:
\begin{align}
  G_{\mu\nu\alpha\beta} = \cfrac{i}{k^2} \left( P^2_{\mu\nu\alpha\beta} -\cfrac12 ~ P^0_{\mu\nu\alpha\beta} \right).
\end{align}
To account for the graviton self-energy the following graviton polarization operator should be used:
\begin{align}
  \Pi_{\mu\nu\alpha\beta} = i \mathcal{N} \kappa^2 k^4 \left[ P^2_{\mu\nu\alpha\beta} + \zeta P^0_{\mu\nu\alpha\beta} \right].
\end{align}
In this expression $\mathcal{N}$ and $\zeta$ are numerical coefficients defined by the structure of the complete quantum model. We do not specify their values, as they are irrelevant for the reasoning to be presented. Moreover, this approach goes in line with classical results \cite{tHooft:1974toh,Goroff:1985th}.

This allows one to recover the resummed graviton propagator:
\begin{align}\label{resummed_propagator}
  \mathcal{G}_{\mu\nu\alpha\beta}=&G_{\mu\nu\alpha\beta} + (G \Pi G)_{\mu\nu\alpha\beta}+\cdots\nonumber\\
  =&\cfrac{i}{k^2} \left(P^2_{\mu\nu\alpha\beta}-\cfrac12~P^0_{\mu\nu\alpha\beta}\right) -\cfrac{i~P^2_{\mu\nu\alpha\beta}}{k^2-\cfrac{-1}{\mathcal{N}\kappa^2}} \\
  & +\cfrac{i~\frac12\,P^0_{\mu\nu\alpha\beta}}{k^2-\cfrac{1}{\frac12 ~\zeta\mathcal{N}\kappa^2}} \nonumber ~.
\end{align}
In full agreement with classical results \cite{tHooft:1974toh,Stelle:1976gc} the propagator contains additional poles corresponding to spin-$0$ massive states and spin-$2$ massive ghosts. These poles mark the applicability limits of the effective theory.

In such a setup an exchange of a virtual graviton between two energy-momentum sources is described by the following expression:
\begin{align}\label{effective_propagator_definition}
  T^{\mu\nu}\mathcal{M}_{\mu\nu\alpha\beta} \left( G^{\alpha\beta\rho\sigma} + \left( G\,  \mathcal{M} \, \Pi \,  \mathcal{M} \,  G \right)^{\alpha\beta\rho\sigma + \cdots}\right) \mathcal{M}_{\rho\sigma\lambda\tau} T^{\lambda\tau}=T^{\mu\nu} \overline{\mathcal{G}}_{\mu\nu\alpha\beta}T^{\mu\nu},
\end{align}
\begin{align}\label{effective_propagator}
  \overline{\mathcal{G}}_{\mu\nu\alpha\beta} =\cfrac{i}{k^2} \left( (1+A)^2~ P^2_{\mu\nu\alpha\beta}-\cfrac12 ~(1+B)^2 ~P^0_{\mu\nu\alpha\beta} \right) \\
  -\cfrac{i~(1+A)^2~P^2_{\mu\nu\alpha\beta}}{k^2 -\cfrac{-1}{\mathcal{N}\kappa^2\,(1+A)^2} } +\cfrac12~\cfrac{i~(1+B)^2~P^0_{\mu\nu\alpha\beta}}{k^2 -\cfrac{1}{\frac12 \zeta\mathcal{N} \kappa^2\,(1+B)^2}}\nonumber ~.
\end{align}
In this expression $\overline{\mathcal{G}}$ is not a resummed graviton propagator, but a quantity that accounts both for graviton propagation and for the mixing. To be short we will call $\overline{\mathcal{G}}$ the effective propagator.

It may seem that effective and resummed propagators have a similar structure, but this is not so. The mixing coefficients $A$ and $B$ are functions of the transferred momentum $k^2$, so they alter the pole structure. For instance, if $A^2$ is proportional to $k^2+1/(\mathcal{N} \kappa^2)$, then the corresponding matrix element is free from the ghost pole. At the same time, if either $A^2$ or $B^2$ admit additional poles, then these poles are to appear in an expression for the matrix element.

The difference between the pole structure of \eqref{resummed_propagator} and \eqref{effective_propagator} shows the following feature of the mixing. The graviton mixing alters the pole structure of the amplitudes involving matter states. This feature has two immediate corollaries.

Firstly, as the graviton mixing influence the pole structure of amplitudes ivolving matter, it also influence the applicability of the effective model. Poles of the resummed graviton propagator \eqref{resummed_propagator} indicate the energy scale at which the effective theory applicability should be put under question, as the ghost instability can be triggered \cite{Solomon:2017nlh}. Poles of the effective propagator \eqref{effective_propagator} play the same role and indicate the limits of the effective theory applicability. Due to the mixing the position of the pole corresponding to ghost states can be altered, which changes the area of applicability for the effective theory.

Secondly, the graviton mixing can be probed empirically. The matrix element of a virtual graviton exchange can be studied empirically, as it defines the form of the Newton law. Namely, it can be studied in the terrestrial environment via experiments of E\"ot-Wash type \cite{Hoyle:2004cw}. The structure of the corresponding matrix element is defined via the effective propagator \eqref{effective_propagator} which contains data on the graviton mixing. Therefore the graviton mixing can be put to a direct empirical verification.

This conclusion is also independently supported with well-known results considering corrections to the Newton potential \cite{BjerrumBohr:2002kt,Bjerrum-Bohr:2016hpa}. In these papers the graviton mixing is not separated explicitly, but it is taken into account. Without the mixing corrections to the Newton potential would have a Yukawa-like form due to the new poles associated with massive states. As the mixing is accounted for, the one-loop effective non-relativistic potential has corrections of a different form.

Finally, one can argue that the graviton mixing is the reason behind the fact the it is impossible to introduce a universal definition of a running gravitational coupling \cite{Anber:2011ut}. As it was highlighted before, within general relativity graviton spin components share the matter coupling. Because of this one can introduce a single coupling constant, the Newton constant, to describe these interactions. When loop corrections are taken into account so the spin components are mixed, the corresponding coupling receive different correction and can no longer be described by a single coupling.

\section{Conclusion}

In this paper we discussed a mechanism of graviton mixing and its possible application within effective field theory for gravity.

We define the graviton mixing in the following way. Within general relativity interaction between matter and spin-2 gravitons (that exist on- and off-shell) alongside with spin-0 gravitons (that exist only off-shell) is defined uniquely. Within effective theory this feature does not hold, as loop corrections can change the corresponding couplings of graviton spin states. We introduce the graviton mixing matrix that accounts for possible mixing of spin-2 and spin-0 graviton states in interaction with an energy-momentum source.

We have shown that graviton mixing strongly influence amplitudes containing matter states. A virtual graviton exchange process was used as an illustrative example. Formulae \eqref{effective_propagator_definition} and \eqref{effective_propagator} show that the poles structure of such a process is defined by the graviton mixing coefficients.

Three conclusions can be made about the graviton mixing. First of all, the mixing can change the area of applicability of the effective theory due to the pole structure of the mixing parameters. Secondly, the mixing can be directly probed empirically. Namely, experiments of E\"ot-Wash type are sensitive to the graviton mixing. Thirdly, the graviton mixing is the reason behind an inability to define a universal running gravitational coupling. The mixing is due loop corrections that has different influence on spin-$2$ and spin-$0$ graviton states coupling to matter. Consequently the corresponding couplings can no longer be described by a single Newton constant.

\bibliographystyle{unsrt}
\bibliography{BibGrph.bib}

\begin{thebibliography}{10}

\bibitem{tHooft:1974toh}
Gerard 't~Hooft and M.~J.~G. Veltman.
\newblock {One loop divergencies in the theory of gravitation}.
\newblock {\em Ann. Inst. H. Poincare Phys. Theor.}, A20:69--94, 1974.

\bibitem{Goroff:1985th}
Marc~H. Goroff and Augusto Sagnotti.
\newblock {The Ultraviolet Behavior of Einstein Gravity}.
\newblock {\em Nucl. Phys.}, B266:709--736, 1986.

\bibitem{Burgess:2003jk}
C.~P. Burgess.
\newblock {Quantum gravity in everyday life: General relativity as an effective
  field theory}.
\newblock {\em Living Rev. Rel.}, 7:5--56, 2004.

\bibitem{Donoghue:2012zc}
John~F. Donoghue.
\newblock {The effective field theory treatment of quantum gravity}.
\newblock {\em AIP Conf. Proc.}, 1483:73--94, 2012.

\bibitem{Donoghue:1994dn}
John~F. Donoghue.
\newblock {General relativity as an effective field theory: The leading quantum
  corrections}.
\newblock {\em Phys. Rev.}, D50:3874--3888, 1994.

\bibitem{BjerrumBohr:2002kt}
N.~E.~J Bjerrum-Bohr, John~F. Donoghue, and Barry~R. Holstein.
\newblock {Quantum gravitational corrections to the nonrelativistic scattering
  potential of two masses}.
\newblock {\em Phys. Rev.}, D67:084033, 2003.
\newblock [Erratum: Phys. Rev.D71,069903(2005)].

\bibitem{Bjerrum-Bohr:2016hpa}
N.~E.~J. Bjerrum-Bohr, John~F. Donoghue, Barry~R. Holstein, Ludovic Plante, and
  Pierre Vanhove.
\newblock {Light-like Scattering in Quantum Gravity}.
\newblock {\em JHEP}, 11:117, 2016.

\bibitem{Calmet:2018qwg}
Xavier Calmet and Boris Latosh.
\newblock {Three Waves for Quantum Gravity}.
\newblock {\em Eur. Phys. J.}, C78(3):205, 2018.

\bibitem{Calmet:2018uub}
Xavier Calmet and Boris Latosh.
\newblock {Dark Matter in Quantum Gravity}.
\newblock {\em Eur. Phys. J.}, C78(6):520, 2018.

\bibitem{Pontecorvo:1957qd}
B.~Pontecorvo.
\newblock {Inverse beta processes and nonconservation of lepton charge}.
\newblock {\em Sov. Phys. JETP}, 7:172--173, 1958.
\newblock [Zh. Eksp. Teor. Fiz.34,247(1957)].

\bibitem{Maki:1962mu}
Ziro Maki, Masami Nakagawa, and Shoichi Sakata.
\newblock {Remarks on the unified model of elementary particles}.
\newblock {\em Prog. Theor. Phys.}, 28:870--880, 1962.

\bibitem{Tanabashi:2018oca}
M.~Tanabashi et~al.
\newblock {Review of Particle Physics}.
\newblock {\em Phys. Rev.}, D98(3):030001, 2018.

\bibitem{Fierz:1939ix}
M.~Fierz and W.~Pauli.
\newblock {On relativistic wave equations for particles of arbitrary spin in an
  electromagnetic field}.
\newblock {\em Proc. Roy. Soc. Lond.}, A173:211--232, 1939.

\bibitem{Hinterbichler:2011tt}
Kurt Hinterbichler.
\newblock {Theoretical Aspects of Massive Gravity}.
\newblock {\em Rev. Mod. Phys.}, 84:671--710, 2012.

\bibitem{RN7}
S.~Weinberg.
\newblock {\em Gravitation and Cosmology : Principles and Applications of the
  General Theory of Relativity}.
\newblock Wiley, 1972.

\bibitem{VanNieuwenhuizen:1973fi}
P.~Van~Nieuwenhuizen.
\newblock On ghost-free tensor lagrangians and linearized gravitation.
\newblock {\em Nucl.Phys.B}, 60:478--492, 1973.

\bibitem{Accioly:2000nm}
A.~Accioly, S.~Ragusa, H.~Mukai, and E.~C. de~Rey~Neto.
\newblock {Algorithm for computing the propagator for higher derivative gravity
  theories}.
\newblock {\em Int. J. Theor. Phys.}, 39:1599--1608, 2000.

\bibitem{Stelle:1976gc}
K.~S. Stelle.
\newblock {Renormalization of Higher Derivative Quantum Gravity}.
\newblock {\em Phys. Rev.}, D16:953--969, 1977.

\bibitem{Solomon:2017nlh}
Adam~R. Solomon and Mark Trodden.
\newblock {Higher-derivative operators and effective field theory for general
  scalar-tensor theories}.
\newblock {\em JCAP}, 1802(02):031, 2018.

\bibitem{Hoyle:2004cw}
C.~D. Hoyle, D.~J. Kapner, Blayne~R. Heckel, E.~G. Adelberger, J.~H. Gundlach,
  U.~Schmidt, and H.~E. Swanson.
\newblock {Sub-millimeter tests of the gravitational inverse-square law}.
\newblock {\em Phys. Rev.}, D70:042004, 2004.

\bibitem{Anber:2011ut}
Mohamed~M. Anber and John~F. Donoghue.
\newblock {On the running of the gravitational constant}.
\newblock {\em Phys. Rev.}, D85:104016, 2012.

\end{thebibliography}

\end{document}